\documentclass[conference]{IEEEtran}
\IEEEoverridecommandlockouts
\PassOptionsToPackage{table}{xcolor}
\usepackage{color,array}

\usepackage{cite}
\usepackage{rotating}
\usepackage[normalem]{ulem}
\useunder{\uline}{\ul}{}
\usepackage[utf8x]{inputenc} 
\usepackage{cite}
\usepackage{amsmath,amssymb,amsfonts}
\usepackage{algorithmic}
\usepackage{graphicx}
\usepackage{textcomp}
\usepackage{xcolor}
\usepackage{array}
\usepackage{longtable}
\usepackage{colortbl}

\def\BibTeX{{\rm B\kern-.05em{\sc i\kern-.025em b}\kern-.08em
    T\kern-.1667em\lower.7ex\hbox{E}\kern-.125emX}}
\begin{document}

\title{A Comprehensive Analytical Review on Cybercrime in West Africa\\
}

\author{\IEEEauthorblockN{Victor Adewopo}
\IEEEauthorblockA{\textit{School of Information Technology} \\
\textit{University of Cincinnati}\\
Cincinnati, USA \\
Adewopva@mail.uc.edu}
\and
\IEEEauthorblockN{Sylvia Worlali Azumah}
\IEEEauthorblockA{\textit{School of Information Technology} \\
\textit{University of Cincinnati}\\
Cincinnati, USA \\
Azumahsw@mail.uc.edu}
\and
\IEEEauthorblockN{ Mustapha Awinsongya Yakubu }
\IEEEauthorblockA{\textit{School of Information Technology} \\
\textit{University of Cincinnati}\\
Cincinnati, USA \\
yakubuma@mail.uc.edu}
 \and
\IEEEauthorblockN{Emmanuel Kojo Gyamfi}
\IEEEauthorblockA{\textit{School of Information Technology} \\
\textit{University of Cincinnati}\\
Cincinnati, USA \\
gyamfieo@mail.uc.edu}
\and
\IEEEauthorblockN{Murat Ozer}
\IEEEauthorblockA{\textit{School of Information Technology} \\
\textit{University of Cincinnati}\\
Cincinnati, USA \\
ozermm@ucmail.uc.edu}
\and
\IEEEauthorblockN{Nelly Elsayed}
\IEEEauthorblockA{\textit{School of Information Technology} \\
\textit{University of Cincinnati}\\
Cincinnati, USA \\
elsayeny@ucmail.uc.edu}

}

\maketitle

\begin{abstract}
Cybercrime is a growing concern in West Africa due to the increasing use of technology and internet penetration in the region. Legal frameworks are essential for guiding the control of cybercrime. However, the implementation proves challenging for law enforcement agencies due to the absence of a dedicated and effective regional institutional follow-up mechanism. This study conducted a systematic literature review focusing on West Africa's prevalence of cybercrime, governing policies, regulations, and methodologies for combating cybercrime. 
West-Africa countries face significant cybercrime challenges, exacerbated by inadequate resources and a dearth of security experts. This study pinpoints potential cybercrime prevention strategies, such as leveraging the Triage framework and broadening research to cover pivotal areas like cyber aggression and cyberbullying. Our research findings highlight the urgency for policymakers and law enforcement agencies to devise more efficient prevention strategies and policies. Overall, this study provides invaluable insights into the state of cybercrime in West Africa to guide the formulation of potent prevention and intervention strategies.
\end{abstract}


\section{Introduction}
Cyberspace holds an enormous quantity of information useful for cybersecurity experts in gathering threat intelligence, preventing cyberattacks, and protecting an organization’s network system \cite{adewopo2022deep}. The landscape of modern technology is rife with security risks due to the extensive network of interconnected devices in the digital sphere. Nonetheless, the pervasive implementation of cyber technologies, the Internet of Things (IoT), and intelligent solutions has been a catalyst for a revolution that has substantially improved the quality of life. Despite its widespread use, the digital domain is one of the most sophisticated systems crafted by humans, comprehensible in its full complexity by only a few \cite{azumah2023cyberbullying}. 

The crypt nature of cyberspace provided a platform for an organized and sophisticated method of cybercrime by cyber actors. Cybercrimes refer to criminal activities conducted via the internet and involve the use of computers either as instruments to commit the crime or as the primary targets of the crime \cite{goyaldigital}. In some cases, both the computer and the person behind it can be victims, depending on the primary target. This means that a computer can either be the target of the attack or the tool used to carry out the crime \cite{goyaldigital}. For instance, hacking involves accessing a computer's information and resources. When the focus of cybercrime is on an individual, the computer is viewed as a tool rather than the target.  
Typically, these crimes do not require an elevated level of technical expertise since they take advantage of human vulnerabilities and result in real-world consequences. The damage caused by such crimes is primarily psychological and intangible, making it challenging to take legal action against those responsible. These types of crimes have existed for centuries, even before the advent of sophisticated technology. Fraudulent activities, theft, and swindles have been common, even without high-tech equipment. However, using computers and the internet has provided criminals with a tool to expand their pool of potential victims and evade detection. As a result, such crimes are committed frequently in the digital age.

The United Nations World Prospects Report of 2022 indicates that West Africa is projected to reach a population of approximately one billion by 2075, with over 500 million individuals falling within the age group of 25 to 64. Additionally, the report highlights a 31.2\% unemployment rate among individuals aged 15 to 24 \cite{UnitedNations}.
Internet access in West Africa exhibits significant variation among its states. Togo stands out with the highest growth rate of 912\% in connectivity between 2000 and 2021, while Cape Verde experienced a more modest growth rate of 4.3\% during the same period. Nigeria leads the region, with over half its population having internet access. In contrast, Niger and Chad face considerable challenges, with internet access rates of 2.2\% and 2.7\% respectively \cite{InCyber}.
Interpol's 2021 report on Cybercrime in Africa as a continent showcases a notable increase in cyberattacks; there was a staggering 238\% surge in cyberattacks targeting online banking platforms. Online scams also pose a significant challenge across Africa; 27\% of web threat detections in Africa in May 2021 were related to online scams. The impact of cybercrime on Africa's economy is substantial, with a reduction of over 10\% in GDP and an estimated cost of 4.12 billion USD in 2021 \cite{INTERPOL}. Trend Micro reported millions of threat detections in Africa between January 2020 and February 2021. These included 679 million email threat detections, 8.2 million file detections, and 14.3 million web detections. South Africa recorded a total of 230 million threat detections, with 219 million of them related to email threats. Kenya and Morocco had 72 million and 71 million threat detections, respectively. Notably, South Africa also had the highest number of targeted ransomware attacks \cite{INTERPOL}.
These disparities in internet access underscore the diverse digital connectivity landscape in West Africa, presenting unique hurdles in addressing cybersecurity and combating cybercrime throughout the region. Moreover, considering the projected population growth and the significant proportion of individuals within the working-age population, it becomes increasingly vital to ensure the development of effective strategies to harness the potential of digital technologies while safeguarding against cyber threats.

Cybercrime in West Africa is unique and different from other parts of the world because there are no adequate tools, resources, and security experts who possess knowledge of combatting cybercrime. Furthermore, the lack of security awareness programs and specialized training for African law enforcement agencies further exacerbates the issue. Many experts are warning that Africa is increasingly becoming a leading source of cybercrime, with Nigeria ranking as the top state in the region for both the target and source of malicious internet activities, and this trend is spreading across the West African sub-region \cite{quarshie2012fighting}.  
It is imperative to research methods for proliferating and preventing the occurrence of cyber-attacks, most especially in the West African region where cyber-attacks are often underreported and lack proper documentation \cite{adewopo2022deep}. An essential aspect of developing effective security measures in the region involves documenting the policies and regulations established by government agencies. By thoroughly examining these policies, researchers and security experts can gain insight into the best practices for protecting against cyber threats and ensure that any security measures put in place are consistent with government regulations. Ultimately, this research can contribute to a safer and more secure cyber environment in West Africa.

\section{History of Cybercrime in West Africa}
In recent years, there has been a significant increase worldwide in the number of incidents related to cybercrime, and West Africa is not an exception \cite{boateng2011sakawa}. According to various reports and analyses, West African countries have experienced a surge in cybercrime activities, with major incidents including phishing scams, ransomware attacks, financial fraud, identity theft, and online scams. These criminal activities have impacted both individuals and businesses and local, regional, and international economies within the sub-region, leading to a significant decrease in trust and confidence in online transactions and financial activities \cite{kshetri2019cybercrime,falowo2022threat}. Compared to previous decades, information and communication technologies (ICTs) have gained significant ground in Africa \cite{boateng2011sakawa}. Although primary access to the internet and other online services in many parts of Sub-Saharan Africa still depends on public access points like cyber cafes, countries such as Nigeria, Cameroon, and Ghana now offer mobile internet access options through fiber optic cables and satellite connections\cite{atta2009understanding}. Computer security experts have found that a significant amount of cybercrime originates from Africa, and this poses a severe threat due to the inadequate protection of computer systems \cite{warner2011understanding}. Unfortunately, poverty and underdevelopment are the primary causes of the growth of cybercrime in the region, making it crucial to have a coordinated approach to combat these threats.  

\subsection{Cybercrime Epidemic in West-Africa}
Cybercrime is more prevalent in  West African countries. Internet users in Ghana reached approximately 17 million, as of January 2022, representing 53\% of the country's population \cite{DataReportal}, while Nigeria is one of the countries mostly affected by cybercrime, according to recent studies \cite{duah2015impact}. Cybercriminals have been able to target people and businesses in Nigeria based on the expansion of digital technology and internet usage, which had severe influence on finances, reputation, and growth of the region \cite{longe2009criminal}. 
Most orchestrated cybercrimes originated locally from Nigeria, but it is noteworthy that these malicious activities involve collaborations with nationals from other countries, such as South Africa and Cameroon \cite{longe2009criminal}. In the 2000s, there was an upsurge in imitations of email scams emerging from various locations in Africa, Asia, and Eastern Europe, which indicates that local elements heavily influence the cybercrime landscape in Nigeria but also have a significant international dimension. Ransomware attacks on Nigerian businesses skyrocketed from 22\% in 2020 to 71\% in 2021, with affected companies paying an average of \$3.43 million compared to \$0.46 million in the previous year. Additionally, external threats are evident with entities like "Anonymous" allegedly targeting major financial infrastructures such as the Central Bank of Nigeria in 2020 \cite{ikusika2022critical}. These cyber vulnerabilities, both domestic and foreign, have not only paralyzed 97\% of affected businesses but also resulted in significant revenue losses for 96\% of them \cite{ikusika2022critical, ede2023impact}. Furthermore, because hackers now operate across borders and jurisdictions, it is challenging for law enforcement organizations to investigate and prosecute cybercrime crimes \cite{boateng2011sakawa}. 
The International Telecommunications Union (ITU) recently released a report that ranked West Africa as having one of the lowest levels of cybersecurity readiness worldwide \cite{duah2015impact}.
The exponential growth in internet usage in West Africa over the past two decades clearly indicates the need for increased attention towards cybercrime prevention in the region. With countries like Togo and São Tomé and Príncipe showing internet usage growth rates of 912 percent and 882 percent, respectively, it is evident that the number of internet users in West Africa is on the rise \cite{InternetUsersWA2023Apr}. Countries such as Cabo Verde, Mali, and Nigeria, with an internet penetration rate above 60\%, are among the top performers in the region. As the number of internet users in West Africa approaches that of the United States, it is imperative that we take heed of the potential risks and consequences of cybercrime. The US, with over 307 million internet users nationwide, has experienced numerous high-profile cyberattacks and severe crimes over the years \cite{InternetUsersWA2023Apr}. It stands to reason that West Africa, with its growing number of internet users, may soon experience similar challenges. As such, investigating and preventing cybercrime in West Africa must be given the attention it deserves.

\begin{figure}[h!]
    \includegraphics[width=0.9\linewidth]{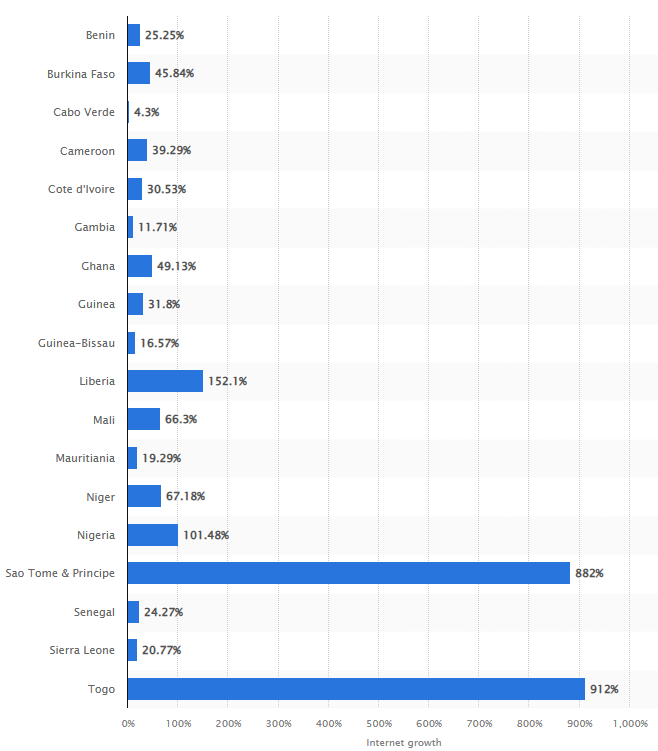}
    \caption{Percentage of Internet Users in West Africa \cite{InternetUsersWA2023Apr}}
    \textit{Alt text: ``Bar chart displaying the percentage distribution of internet users in West Africa. Togo has a notably high 912\% users, followed by Sao Tome and Principe with 882\%, and Liberia with 152.1\%. The percentages represent the proportion of the population in each country that uses the internet."}
    \label{internetusers}
\end{figure}

 \subsection{Landscape of Cybercrime} \label{strategy}

The high rate of cybercrime in the area is significantly influenced by a lack of preparedness and sophisticated cyber technologies. Anderson et al. \cite{anderson2013measuring}, reported 33\% increase in instances of cybercrime reported in West-Africa. A growing number of people and businesses are becoming victims of online fraud, identity theft, and other types of cybercrime in West Africa, which is a cause for concern. Cybercrime has advanced and grown more complex in recent years, providing considerable difficulties for regional governments and law enforcement organizations. It is crucial to comprehend the nature, scope, and trends of cybercrime in West Africa in order to address it effectively. Cybercrime has increased globally as a result of more people using the internet and other technology. With a growth in cybercrime activities over the past few years, West Africa has been included in this trend. Cybercrime in West Africa is a prominent issue that has dire consequences on the quality of human lives, businesses, governments, and growth of West Africa \cite{duah2015impact}.

In the domain of cybersecurity, the dynamic interplay between attackers and defenders can be aptly likened to a game of cat and mouse. Attackers persistently scour for system vulnerabilities, while defenders diligently strive to identify and remediate these weaknesses preemptively. Researchers have turned to game theory as a valuable tool within the realm of cybersecurity \cite{shiva2012holistic, lye2005game}. By providing a structured framework, game theory facilitates the comprehension of strategic interactions between attackers and defenders, offering insights into decision-making processes where outcomes are contingent upon the actions of multiple players. Shiva et al. \cite{shiva2012holistic}, conducted research that applied game theory to model a general sum Nash equilibrium specifically in the context of cybersecurity. This pioneering work enabled system administrators to gain crucial insights into the strategies and plans of attackers. By leveraging game theory, defenders are empowered to develop effective defense strategies against potential and active cyber threats.

This method allows security experts to capture the probabilistic nature of transition, which is vital in the constantly evolving world of cyber threats. One limitation of this approach is the large state space where multiple endpoints can be attacked, making it challenging to manage. Additionally, administrators only need to act when an attack is suspected, complicating the decision-making process. Balancing security measures and usability also presents challenges, as overly restrictive policies may hinder users' productivity. To address these concerns, Shiva et al.\cite{shiva2012holistic} research utilized gametheory approach in analysis of interactions between attackers and system administrators (security specialists). As the cybersecurity landscape continues to evolve, game theory proves indispensable in understanding and mitigating the intricate dynamics between attackers and defenders.

\subsection{Legal Framework and Policies Governing Cybercrime}
It is imperative that governments and organizations within West Africa invest in improving their cybersecurity measures, implementing effective policies and regulations, and enhancing public awareness. Some West African countries have taken steps to address the rising trend of cybercrime by strengthening their security infrastructure, collaborating with international organizations, law enforcement agencies, conducting awareness campaigns, and making cybersecurity education and training more accessible to citizens \cite{orji2015multilateral}. 

The pattern of cybercrime in West Africa has been explored by other researchers. A study conducted by Adomi et al. \cite{adomi2008combating} looked into the frequency of cybercrime in Nigeria as well as the various forms it can take. According to the findings of the study, the three types of cybercrime that occur the most frequently in the country are phishing, social engineering, and identity theft. Another study by Frank et al. \cite{frank2013approach} investigated the effects that cybercrime had on Nigerian companies \cite{frank2013approach}. According to the survey findings, businesses are losing millions of dollars annually due to the activities of cybercriminals, which has a major and detrimental effect on the economy. Based on the proliferation of online criminal activity in West Africa, more and more people are working to put a stop to it. In an effort to prevent cybercrime and improve cybersecurity readiness, a number of projects have been initiated by various groups and governments. For instance, in 2018, the West African Cybersecurity and Cybercrime Conference (WACCC) was established to bring together various stakeholders in the region in order to debate various problems and potential solutions pertaining to cybersecurity \cite{abubakari2021reasons,duah2015impact}. Since then, the conference has become an annual event, drawing attendees from West Africa and further afield. 

In Nigeria, cybercrime has progressed from "yahoo yahoo" to "yahoo plus," which indicates a high level of progression in illegal operations that law enforcement agencies and victims have not yet come to terms with \cite{Akanle2020Mar}.
In nations across West Africa, including Nigeria and Ghana, a variety of legislations targeting cybercrime have been put into effect \cite{kolog2023imple}. Such legal structures are crucial in guiding the regulation of cyber activities. Nevertheless, the practical application of these legal systems presents complexities and hurdles for enforcement bodies. The enactment of the ECOWAS 2011 directive within this region, for instance, has encountered notable obstacles, as detailed by Orji \cite{JeromeOrji2019Nov}. A key hindrance is the lack of a robust regional mechanism dedicated to the effective enforcement of the directive, which would support the pursuit of its aims across the member states. Compounding the problem is the deficit of skilled professionals capable of assisting in the crafting and enforcing of cybercrime-related legal, regulatory, and procedural frameworks. Moreover, the sluggish evolution of legal frameworks in many emerging economies further impedes the swift implementation of such cybercrime initiatives. Therefore, the implementation of legal frameworks to control cybercrime requires a well-structured institutional follow-up mechanism, expert personnel, and a streamlined legal initiative development process.
The study of Kademi et al. \cite{kademi2018strengthening} identifies and discusses crucial concerns within the Nigerian National Cyber Security Strategy (NNCSS) and proposes measures to strengthen it. In order to cater to the Nigerian context and include key aspects of strategy development, they proposed a framework model to enhance the strategy. To strengthen the country's cyber security efforts, the authors recommend enhancing institutional adjustments and fostering collaboration across levels.
The study also provides several suggestions to strengthen the National Cyber Security Strategy (NNCSS) against cybercrime. These suggestions include:
\begin{enumerate}
    \item \textit{Developing and implementing performance measures: }The paper suggests that NNCSS should have a set of milestones and timeframes to track progress in achieving its objectives. Developing and implementing performance measures is essential in evaluating the effectiveness of action controls.
    \item \textit{Integrating cyber security exercise with capacity building process: }The paper recommends that a plan to integrate cyber security exercise with capacity building process should be encouraged. This will help to enhance the capacity of stakeholders to respond to cyber threats.
    \item \textit{Giving more priority to research and development: }The paper suggests that research and development should be given more priority as all routes to cyber capacity development must be explored.
    \item Adjusting the institutional setting: The paper highlights the need for an adjustment to the institutional setting to strengthen the country's cyber security efforts.
    \item \textit{Digital forensics / Cyber threat Intelligence: }The establishment of a digital forensic laboratory and capacity building is necessary to combat persistent and sophisticated cyber intrusions and help in the fight against insurgency and crimes. (Context: Cyber espionage necessitates the development of counter-intelligence measures and foreign policy response.)
    \item  \textit{Partner with ISP: }Collaboration with internet service providers and other organizations can help identify and respond to cyber threats while maintaining privacy.
\end{enumerate}
This emphasizes the significance of developing a tailored and comprehensive cybersecurity strategy for West African countries, taking into account the unique challenges and potential opportunities within the region.


\section{Methodology}\label{Methods}
The literature search process conducted in this investigation entailed a four-step method: i) determining the inclusion and exclusion criteria for eligibility; ii) outlining the research objectives; iii) crafting a search strategy; and iv) extracting the data~\cite{sarkis2021properly}.
In this study, a systematic review methodology was utilized to systematically and reproducibly answer the research questions presented. In particular, the PRISMA (Preferred Reporting Items for Systematic Reviews and Meta-Analyses) Statement provided the methodological framework for this review~\cite{sarkis2021properly}. The papers that met the pre-established eligibility criteria were meticulously analyzed and synthesized to address the research questions laid out in the subsequent subsection.

\subsection{Research Questions and Objectives}\label{Rqs}
Cybercrime is a major societal issue that involves multiple actors scouring for security weaknesses to exploit. Cybersecurity experts are focused on the defensive, seeking innovative ways to employ emerging technologies to safeguard organizational assets and data \cite{goni2022basic}. Some key considerations for thwarting cyber-attacks include identifying high-risk systems, understanding the tools utilized by cyber criminals, and determining the potential impact on compromised systems. Cybercrime in West Africa is differentiated into three life cycles, including Scouting and harvesting, relationship building, and operational stages \cite{atta2009understanding}. 

This study implores a systematic literature review methodology to understand the current state of cybercrime in West Africa, including the latest trends and patterns, types of attacks, most vulnerable sectors and populations, and impacts on individuals, businesses, and the economy as a whole, cybersecurity measures currently in place, policy and regulatory landscape, ongoing initiatives to combat cybercrime across the sub-region. 

This systematic literature review's objective is to assess the present state of knowledge regarding cybercrime in West Africa. In particular, this analysis will look at the frequency and forms of cybercrime in the area, their effects on people, businesses, and governments, and the steps taken to counteract it. 
In this systematic literature review, we seek to find answers to the following questions. 
\begin{itemize}
    \item \textbf{\textit{RQ1:}} What are the current techniques and methodology for combating cybercrime?
    
    \item \textbf{\textit{RQ2:}} What are the potential strategies for effective cybercrime prevention in West Africa?
\end{itemize}

\subsection{Selecting Eligibility Criteria}
The scope of this review encompasses scholarly articles addressing Cybercrime within the West African region. Covered subjects include the incidence of Cybercrime in West Africa, approaches to prevention, and strategic proposals for mitigating cybercrime, with sources restricted to those featured in peer-reviewed publications from 2013 to 2023. Reflecting the rapid pace of technological progress, the articles included in this review span a period of ten years prior to this study. The review is limited to studies published in English. For clarity, the criteria guiding the selection of articles are explicated in Subsection~\ref{inclusion}, while the grounds for excluding studies are specified in Subsection~\ref{exclusion}.

\subsection{Inclusion Criteria} \label{inclusion}
To be considered for inclusion, the publications were required to possess the following characteristics:
\begin{enumerate}
\item Articles should be in Cybercrime domain.
\item Published within 2013-2023 (10 Years).
\item Analyzed /discussed cybercrime data.
\item Study conducted in West Africa.
\item Full /short papers that are peer-reviewed.
\end{enumerate}

\subsection{Exclusion Criteria}\label{exclusion}
The following exclusions were implemented:
\begin{enumerate}
\item Does not contain Cybercrime analysis. 
\item Published before 2013. 
\item Not peer reviewed or does not provide clear findings and analysis of results.
\item Written in other languages excluding English. 
\item Duplicated studies.
\end{enumerate}

\subsection{Information Sources}
\begin{figure}[h!]
\centering
\includegraphics[width=0.9\linewidth]{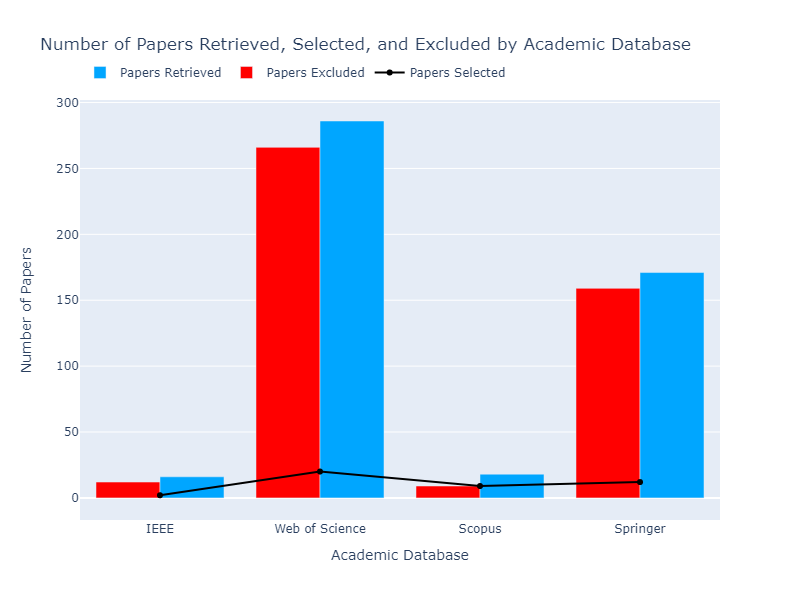}
\caption{Proportion of selected studies}
\textit{Alt text: “Figure 2 shows the number of articles retrieved from each database and the final number of selected papers.”}
\label{selected}
\end{figure}

\begin{table}
	\small\addtolength{\tabcolsep}{-5pt}
	\caption{Article Data source}
	\label{database}
	\footnotesize
	\begin{tabular}{l|l|l|ll}
		\hline
		\textbf{ID} & \textbf{Database}                      & \textbf{Link} & \textbf{Number of Articles}& \\ \hline
		D1          & \cellcolor[HTML]{C0C0C0}IEEE Xplore    & { http://ieeexplore.ieee.org/} & 16 &  \\ \hline
		D2          & \cellcolor[HTML]{C0C0C0}Scopus            & { https://www.scopus.com/home.uri/} & 18 &  \\ \hline
		D3          & \cellcolor[HTML]{C0C0C0}Web of Science & {
			http://www.webofscience.com/} & 286 &  \\ \hline
		D4          & \cellcolor[HTML]{C0C0C0}Springer Link & { http://link.springer.com/} &  171&  \\ \hline

	\end{tabular}
\end{table}
The selection of articles for inclusion in this review was accomplished through a strategic search of electronic databases featuring publications in English.
Table~\ref{database} details the databases that served as the foundational sources for literature relevant to cybercrime in West Africa, encompassing significant full-text scholarly journals and conference proceedings.
The preliminary step entailed an advanced search within the enumerated databases in Table~\ref{database}, applying rigorous filtering criteria to distill the results to those strictly applicable to the study area. In the ensuing phase, two research assistants rigorously assessed the search outputs manually to ascertain the relevance and integrity of the findings. The volume of articles retrieved from each listed database, as well as the final articles that were eventually chosen for review, is presented in Figure~\ref{selected}. The inclusion was restricted to articles that were available for review. Further explication of the search methodology and the criteria for validating and electing the appropriate articles are elaborated in Subsection~\ref{strategy}.

\subsection{Search Strategy} \label{strategy}
Combining the following keywords with conjunctions ``AND" and disjunctions ``OR" resulted in a total of 490 papers in an automated search, as shown in Table~\ref{database}. The most common terms used for our search were:
\begin{enumerate}
	\item Cybercrime.
	\item Fraud.
	\item Sakawa.
	\item Gameboys.
        \item Money and Theft.
        \item Yahoo Yahoo.
        \item 419.
\end{enumerate}

\noindent
The results of our search and the corresponding query that has been used are as follows:
\begin{itemize}
\item {\textbf{\textit{IEEE Xplore:}}} We received~\textit{16 papers} from IEEE using the search string: \texttt{[("Full Text Only":Cybercrime in WEST AFRICA) AND ("All Metadata":Fraud) OR ("Full Text Only":Cybercrime in WEST AFRICA) AND ("All Metadata":Yahoo Yahoo) OR ("Full Text Only":Cybercrime in WEST AFRICA) AND ("All Metadata":Internet Theft) OR ("Full Text Only":Cybercrime in WEST AFRICA) AND ("All Metadata":Game boys) OR ("Full Text Only":Cybercrime in WEST AFRICA) AND ("All Metadata":sakawa OR 419) OR ("All Metadata":Mobile Money Theft)] between 2013-2022}

 \item {\textbf{\textit{Web of Science:}}} We received 286 papers from Web of Science using the search string: \texttt{[((ALL=(Cybercrime) AND ALL=(west africa OR Fraud OR sakawa OR 419 OR yahoo yahoo))) AND (PY==("2023" OR "2022" OR "2021" OR "2020" OR "2019" OR "2018" OR "2017" OR "2016" OR "2015" OR "2014" OR "2013" ))]}
	
\item {\textbf{\textit{Springer Link:}}} We received \textit{171 papers} from Springer Link using the search string: \texttt{[Cybercrime AND "africa" AND (Fraud, OR sakawa, OR gameboys,419, OR money OR theft, OR internet OR scam, OR yahoo OR yahoo, OR gameboys)] between 2013-2022 }
	
\item {\textbf{\textit{Scopus:}}} We received \textit{18 papers} from Science Direct using the search string: \texttt{[((TITLE-ABS-KEY(cybercrime) AND TITLE-ABS-KEY(fraud) OR TITLE-ABS-KEY(sakawa) OR TITLE-ABS-KEY(gameboys) OR TITLE-ABS-KEY(419) OR TITLE-ABS-KEY(yahoo AND yahoo) OR TITLE-ABS-KEY(internet AND scam) OR TITLE-ABS-KEY(money AND theft) AND TITLE-ABS-KEY(africa))] between 2013-2022 }

\end{itemize}

\subsection{Study Selection}

\begin{figure}[h!]
	\centering
	\includegraphics[width=1\linewidth,height=1\linewidth,keepaspectratio]{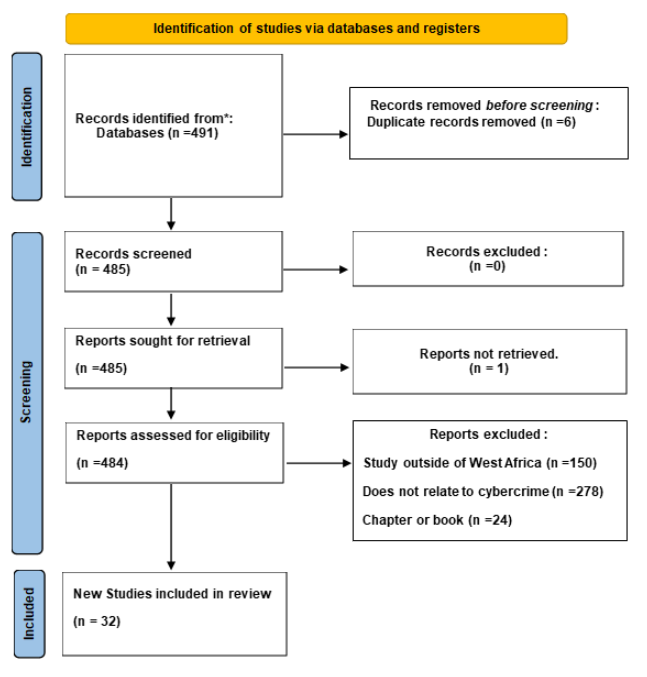}
	\caption{Preferred Reporting Items for Systematic Reviews and Meta-Analyses (PRISMA) flow chart of the systematic review.}
\textit{Alt text:``Figure 3 presents a PRISMA flow chart illustrating the systematic review process undertaken. It depicts the initial identification and screening of 491 articles, detailing exclusions and inclusions due to relevance to crime analysis or policies in West Africa."}
	\label{Reference: Methodology}
\end{figure}

The study incorporated articles that were meticulously chosen based on the established selection guidelines as presented in Section (\ref{Methods}). The initial search through designated databases, guided by a specific search strategy, identified 491 potential articles. These were subsequently examined for their eligibility by a team of four student researchers with expertise in the pertinent field. An independent review of the abstract, title, and keywords of each paper was carried out to ensure its relevance to the posed research question and adherence to the inclusion and exclusion criteria.
Non-qualifying articles, or those unrelated to the study's scope, were systematically excluded. The student researchers independently evaluated and scored each article according to the set eligibility requirements. This selection protocol guaranteed the relevance of the articles ultimately included in the research.
From the initial pool, 448 articles were excluded for not addressing the specific topics of crime analysis or crime-related policy in the West African context. A further six were removed due to being duplicates. Consequently, 32 papers met the criteria and were chosen for in-depth review, as depicted in Figure~\ref{Reference: Methodology}. 
\begin{table*}
\centering
\caption{Summary of research articles selected, the main theme presented and research question addressed. The notation ``-" implies the research article does not address the research question.}
\renewcommand*{\arraystretch}{1.8}
\label{tab:my-table3}
\begin{tabular}{|c|c|c|c|c|}
\hline
\rowcolor[rgb]{0.678,0.678,0.678} \textbf{Author} & \textbf{Main Theme} & \textbf{RQ1} & \textbf{RQ2} & \textbf{Prevention Method}\\
\hline
Ibrahim et al. \cite{Ibrahim2016Dec} &
  Cybersecurity &
  \checkmark &
  - &
  Economic austerity measures \\ \hline
Mba et al. \cite{Mba2017Apr} &
  Cybercrime &
  \checkmark &
  - &
  ML Algorithms \\ \hline
Orji et al. \cite{JeromeOrji2019Nov} &
  Legal Frameworks &
  \checkmark &
  - &
  ECOWAS Community Laws \\ \hline
Ogunleye et al. \cite{Ogunleye2020Mar} &
  Cybercrime &
  \checkmark &
  - &
  Whistleblowing, Education, Law Enforcement \\ \hline
Tade et al. \cite{tade2019cybercrime} &
  Cybercrime &
  - &
  - &
  - \\ \hline
Tade et al. \cite{Tade2020Jun} &
  Financial Crime &
  - &
  \checkmark &
  Short Messaging Services (SMS) \\ \hline
Lazarus et al. \cite{Lazarus2019Jul}. &
  Financial Crime &
  - &
  - &
  - \\ \hline
Genc et al. \cite{Genc2021Mar} &
  Cybercrime &
  \checkmark &
  - &
  Austerity measures to the ailing economy and creating jobs for the teaming youths. \\ \hline
Olayinka et al. \cite{Akanle2020Mar} &
  Cybersecurity &
  - &
  - &
  ML Algorithms \\ \hline
Lazarus et al. \cite{Lazarus2022Nov} &
  Cybercrime &
  - &
  - &
  - \\ \hline
Alhassan et al. \cite{Alhassan2021May} &
  Cybersecurity &
  - &
  - &
  Awareness, Cultural and social factors \\ \hline
Abayomi-Alli et al. \cite{Abayomi-Alli2022Mar} &
  Cybercrime &
  - &
  - &
  - \\ \hline
Akanle et al. \cite{Akanle2020Apr} &
  Financial Crime &
  \checkmark &
  \checkmark &
  Collaboration, Government, Civil Society \\ \hline
Beek et al. \cite{Beek2016May} &
  Cybercrime &
  - &
  - &
  Awareness, Social factors, Story telling \\ \hline
Okpa et al. \cite{Okpa2022Jun} &
  Financial Crime &
  \checkmark &
  \checkmark &
  Training, Awareness \\ \hline
Kademi et al. \cite{kademi2018strengthening} &
  Legal Frameworks &
  \checkmark &
  \checkmark &
  Strategic decisions /regulations \\ \hline
Ukeachusim et al. \cite{ukeachusim20221} &
  Cybersecurity &
  - &
  - &
  Biblical methods \\ \hline
Olofinbiyi et al. \cite{olofinbiyi2021exploring} &
  Cybercrime &
  - &
  \checkmark &
  Cybercrime readiness \\ \hline
Hamisu et al. \cite{hamisu2021analysis} &
  Legal Frameworks &
  \checkmark &
  - &
  Strategic recommendation \\ \hline
Ojukwu-Ogba et al. \cite{ojukwu2020legal} &
  Legal Frameworks &
  - &
  - &
  Legal prevention \\ \hline
Omodunbi et al. \cite{omodunbi2016cybercrimes} &
  Cybersecurity &
  \checkmark &
  \checkmark &
  - \\ \hline
Oludayo Tade \cite{tade2013spiritual} &
  Cybercrime &
  - &
  - &
  - \\ \hline
Kanu et al. \cite{kanu2022frauds} &
  Financial Crime &
  - &
  \checkmark &
  Block chain technology, training on e-banking \\ \hline
Isacenkova et al. \cite{isacenkova2014inside} &
  Cybercrime &
  \checkmark &
  \checkmark &
  TRIAGE Framework, Analysis of scam patterns \\ \hline
Ephraim et al. \cite{ephraim2013african} &
  Cybercrime &
  \checkmark &
  \checkmark &
  Culture-centered Approach , Government and religious institution \\ \hline
David-West et al. \cite{David-West2021Sep} &
  Financial Crime &
  - &
  - &
  - \\ \hline
Abdulrauf et al. \cite{abdulrauf2017personal} &
  Legal Frameworks &
  - &
  \checkmark &
  Legislation and Government regulation \\ \hline
Adeleke et al. \cite{Adeleke2022Jun} &
  Financial Crime &
  - &
  \checkmark &
  - \\ \hline
Dziwornu et al. \cite{dziwornu2021crime} &
  Cybercrime &
  - &
  \checkmark &
  Analysis of crime pattern \\ \hline
Olubusoye et al. \cite{Olubusoye2023Apr} &
  Cybercrime &
  - &
  - &
  - \\ \hline
Olanrewaju Lawal \cite{Lawal2022Nov} &
  Cybercrime &
  - &
  - &
  - \\ \hline
\end{tabular}
\end{table*}
\section{Results}
\subsection{RQ1 - What are the current techniques and methodology
for combating cybercrime?}
The data presented in table (\ref{tab:my-table3}) underscore the key themes and prevention methods discussed in the selected papers. One notable approach is the method proposed by Isacenkova et al. \cite{isacenkova2014inside}, which emphasizes linking scam activities through consistent contact points such as phone numbers and email addresses. These elements are critical as they typically remain unchanged over extended periods, thus acting as vital links between the perpetrators and their victims. The experiments conducted in the study demonstrated that this approach can reliably trace macro clusters of scam activities and identify new scam operations by associating ongoing campaigns with new initiatives. The research analyzed three years of data obtained from a well-known scam aggregator website, "419Scam.org" which provides information on spam activities. The study also investigated the use of phone numbers to tag the geographical location of where spam originates. However, this method has its limitations, as it does not conclusively prove the origin of such scam messages. Over the three-year period, the data revealed that Nigeria, the United Kingdom, and Benin had the highest incidence of mobile phone number scams related to 419 activities~\cite{isacenkova2014inside}. Ephraim et al.~\cite{ephraim2013african} introduced a culture-centered approach, utilizing fundamental principles to guide the responsible use of cyberspace and prevent crimes such as bullying, extortion, and violence. Culture aims to promote norms that establish the duty to abstain from criminal activities; a culture-centered strategy encompasses four primary components: 1) educational institutions and government agencies, 2) families, 3) religious organizations, and 4) mass media~\cite{ephraim2013african}. By incorporating these elements, a proactive culture-centered approach can foster a sense of responsibility and awareness among individuals and help to create a safe online environment, empowering communities to protect themselves and others from various forms of cybercrime.
With most cybercrimes now perpetuated online, Mba et al. \cite{Mba2017Apr} devoted two years to understudying an online forum used by scammers to advertise various schemes aimed at deceiving and exploiting their victims. Their study proposed leveraging Machine Learning (ML) Algorithms and techniques to examine huge datasets like the one they used for their study. Such ML algorithms not only speed up the data analysis process but also accurately predict successful cybercrime transactions executed on some of the fraudulent websites and portals.
Ogunleye et al. \cite{Ogunleye2020Mar}, on the other hand, in their study investigated the motivation factors of the involvement of female undergraduate students in cybercrime from selected universities across Nigeria. The findings revealed that peer pressure and financial struggles were the primary factors driving this involvement, while some participants also cited "fun and entertainment" as another contributing factor. In their study \cite{Ogunleye2020Mar}, the authors proposed a multi-level approach to mitigate these incidences, including increased engagement of relevant security organizations, such as the Economic and Financial Crimes Commission (EFCC), the establishment of a whistle-blowing reward system, and the expansion of educational programs to raise awareness on the consequences of cybercrime throughout Nigeria.

Whilst Ogulenye et al. \cite{Ogunleye2020Mar} enumerated the key factors influencing the teaming youth, especially young females in universities, the Economic Community of West African States (ECOWAS) in 2011 introduced a legal framework to tackle cybercrime within the sub-region. Orji \cite{JeromeOrji2019Nov} analyzed this particular ECOWAS directive in a 2019 study, examining its adoption by certain member states and discussing the implications of the legal implementation of the Directive's obligations in those countries.
The study by Omodunbi et al.\cite{omodunbi2016cybercrimes} focused on utilizing machine learning algorithms for detecting cyber attacks. The authors highlight the ability of these algorithms to analyze network traffic and detect suspicious patterns, including known attack signatures and anomalies in network behavior. The significance of collaboration between law enforcement agencies and private sector organizations in detecting and combating cybercrime is also emphasized. The authors suggest adopting proactive measures such as implementing security protocols and educating users to prevent cybercrime.\\
Akanle et al. \cite{Akanle2020Apr} recommend a multi-stakeholder strategy for cybercrime in Nigeria that includes government agencies, commercial sector organizations, civil society groups, and international partners. Strategies suggested include Creating and executing cybercrime laws, building law enforcement agencies' technical capacity to investigate and prosecute cybercriminals, and increasing cybercrime and cybersecurity awareness and education. Furthermore, promoting information sharing and collaboration among law enforcement, internet service providers, and financial institutions was encouraged while also highlighting that Nigeria's cybercrime problems require a comprehensive and coordinated strategy.

\subsection{RQ2 - What are the potential strategies for effective cybercrime prevention in West Africa?}

Financial fraud in Nigeria has been on the rise, with Automated Teller Machine (ATM) fraud being one of the most used avenues by these perpetrators \cite{Tade2020Jun}. In their study, Tade et al. \cite{Tade2020Jun} discussed how banks in Nigeria were using Short Messaging Services (SMS) Alert Systems to constantly communicate with their customers about every known cybercrime information that comes to the notice of institutions.
\begin{table*}
\centering
\footnotesize 
\caption{Key Summary of research articles selected.}
\renewcommand*{\arraystretch}{0.8}
\label{summarytable}
\begin{tabular}{|p{0.10\textwidth}|p{0.89\textwidth}|}
\hline
\rowcolor[rgb]{0.678,0.678,0.678} 
\textbf{Author} & \textbf{Key Summary} \\
                   
\hline
Ibrahim et al. \cite{Ibrahim2016Dec}           & The paper aimed to establish particularities of cybercrime in Nigeria and whether they suggest problems with prevailing taxonomies of cybercrime. By doing so, they are able to ensure cybercrimes are categorized accordingly in their rightful groups. The paper also argued that cybercrime was motivated in three possible ways: socioeconomic, psychosocial, and geopolitical.                                                                                                                                                                                                                                                                                                                                                                                                                                                                                                                                      \\ 
\hline
Mba et al. \cite{Mba2017Apr}                   & The paper implored ML algorithms to identify cybercrime perpetuated by Nigerians living in Nigeria from a dataset of an online forum (a crime hub used by scammers to advertise various forms of schemes aimed at deceiving and exploiting their victims). Their study was able to establish the modus operandi of these cyber criminals and their targeted groups of mostly vulnerable society.                                                                                                                                                                                                                                                                                                                                                                                                                                                                                                                          \\ 
\hline
Orji et al. \cite{JeromeOrji2019Nov}           & The study examined the ECOWAS directive's applicability in member states, highlighting legal implementation implications and challenges. Obstacles included the lack of effective regional follow-up mechanisms, expert personnel, and slow-paced legal initiatives in developing countries. \\ 
\hline
Ogunleye et al. \cite{Ogunleye2020Mar}         & This study examined the motivations of cybercrime involvement among female undergraduate students in some selected universities in Nigeria. They found that financial obligations and peer pressure were among some of the main motivations.                                                                                                                                                                                                                                                                                                                                                                                                                                                                                                                                                                                                                                                                              \\ 
\hline
Tade et al. \cite{tade2019cybercrime}          & The paper examined how the lyrics of Nigerian songs were prolificated with cyber criminality.                                                                                                                                                                                                                                                                                                                                                                                                                                                                                                                                                                                                                                                                                                                                                                                                                             \\ 
\hline
Tade et al. \cite{Tade2020Jun}                 & This paper investigated financial crimes and emphasized the numerous innovations banking institutions employed in dealing with cybercrime, including SMS, etc.                                                                                                                                                                                                                                                                                                                                                                                                                                                                                                                                                                                                                                                                                                                                                            \\ 
\hline
Lazarus et al. \cite{Lazarus2019Jul}.            & This study explored the characteristics of Nigerian cybercriminals. It should examine the perspectives of the authority at the forefront of handling cybercrime in Nigeria, thus the Economic and Financial Crimes Commission (EFCC).                                                                                                                                                                                                                                                                                                                                                                                                                                                                                                                                                                                                                                                                                           \\ 
\hline
Genc et al. \cite{Genc2021Mar}                 & The study examined how scammers utilized emails in scamming their victims. The authors implored machine-learning techniques to detect scams in email content.                                                                                                                                                                                                                                                                                                                                                                                                                                                                                                                                                                                                                                                                                                                                                             \\ 
\hline
Olayinka et al. \cite{Akanle2020Mar}            & The study described how cybercrime has evolved from "Yahoo Yahoo" to "Yahoo Plus," indicating a significant advancement in illicit activities that outpaces law enforcement and victim understanding. The authors indicated that the dynamic and ever-evolving issue of cybercrime requires further research to comprehend its complexities and manifestations in Nigeria and West African countries.                                              \\ 
\hline
Lazarus et al. \cite{Lazarus2022Nov}           & Twitter users use linguistic techniques for criticizing Nigerian politicians, and hackers are examined in the article. The Economic and Financial Crimes Commission (EFCC) tweets are examined to determine public opinion on Nigerian cybercriminal charges. The article shows that Twitter users use moral disengagement to justify cyber criminals' actions and defend them by comparing them to corrupt politicians. The study concludes that morality is involved in classifying some activities as crimes and others as not.                                                                                                                                                                                                                                                                                                                                                                                       \\ 
\hline
Alhassan et al. \cite{Alhassan2021May}         & This research examines how Sakawa boys from northern Ghana utilize their bodies and cultural objects to express themselves online. The study found that SSakawa boys" reveal who they are and figure out what they mean by employing slang and jargon, conspicuous use of material items, an extravagant lifestyle, techno religiosity, and cooperating with girls to conduct cyber-fraud. The report discusses the cultural and socioeconomic elements that make Ghanaian youth steal. The study also reveals how crucial education and awareness-raising are for preventing vulnerable youth from becoming cybercriminals.                                                                                                                                                                                                                                                                                            \\ 
\hline
Abayomi-Alli et al. \cite{Abayomi-Alli2022Mar} & The study delves into cybercrime in Nigeria by analyzing public sentiment through the examination of "yahoo-yahoo" related tweets. Utilizing a corpus of 5500 tweets, it employed sentiment analysis to gauge public perceptions of this pervasive form of cybercrime.                                                                                                                                                                                                                                                                                                                                                                                                                                                                                                                                    \\ 
\hline
Akanle et al. \cite{Akanle2020Apr}             & The research shows Nigeria's cybercrime prevention techniques are inadequate. The authors suggested a multi-stakeholder approach involving the government, the commercial sector, civil society, and international partners to combat cybercrime. This approach should include developing and implementing effective legal frameworks and policies, building law enforcement agency technical capacity, and increasing public awareness.                                                                                                                                                                                                                                                                                                \\ 
\hline
Beek et al. \cite{Beek2016May}                 & This research examines Ghanaian email scams and storytelling. It claims that each cybercrime case is a collection of interconnected stories and that scammers, victims, and police personnel are affected by cultural imaginaries and practices. The study discusses global cybercrime prevention and indicates that local cultural differences may need to be considered. The research sheds light on West African cybercrime's complicated dynamics and storytelling's influence.                                                                                                                                                                                                                                                                                                                                                                                                                                       \\ 
\hline
Okpa et al. \cite{Okpa2022Jun}                 & The article investigates the Business Email Compromise (BEC) scam's impact on corporate economic stability in Cross River State, Nigeria, noting its increasing incidence in key sectors like banking and telecommunications. The authors recommend a joint cybersecurity initiative between businesses and government to protect infrastructure, suggesting robust access control and enhanced cyberspace monitoring to aid government anti-cybercrime measures.                                                                                                                                                                                                                                                         \\ 
\hline
Kademi et al. \cite{kademi2018strengthening}   & This paper identifies crucial concerns within the Nigerian National Cyber Security Strategy (NNCSS) and suggests ways to bolster it.                                                                                                                                                                                                                                                                                                                                                                                                                                                                                                                                                                                                                                                                                                                                                          \\ 
\hline
Olofinbiyi et al. \cite{olofinbiyi2021exploring} & The study examines the rise of cybercrime among Nigerian youth, attributing it to factors such as urbanization and unemployment. It also reveals a dichotomy: while some youths distance themselves from cybercrime, others justify it as a response to socioeconomic adversity.                                                                                                                                                                                                                                                                                                                                                                                                                                                                                                                                                                                                        \\ 
\hline
Hamisu et al. \cite{hamisu2021analysis}        & This paper presents a comprehensive outline of cybercrime in Nigeria, and$\sim$provides recommendations and suggestions for enhancing the effectiveness of law enforcement and government efforts to combat cybercrime in Nigeria.                                                                                                                                                                            \\ 
\hline    
Omodunbi et al. \cite{omodunbi2016cybercrimes} & The paper focuses on the utilization of machine learning algorithms for detecting cyber attacks.                                                                                                                                                                                                                                                                                                             \\ 
\hline
Oludayo Tade \cite{tade2013spiritual}          & This paper examines the cybercrime strategy known as Yahoo Plus, which combines spiritual elements with internet browsing to enhance victimisation on the web.                                                                                                                                                                                                                                                                                                                                                                                                                                                                                                                                                                                                                                                                                                                                                                                                                                                                        \\ 
\hline
Kanu et al. \cite{kanu2022frauds}              & This study explored current fraud and forgery cases in the Nigerian banking industry.                                                                                                                                                                                                                                                                                                                                                                                                                                                                                                                                                                                                                                                                                                                                                                                                                                     \\ 
\hline
Isacenkova et al. \cite{isacenkova2014inside}  &  This study investigated various scam campaigns, their methodologies, and geographical distribution while demonstrating the effectiveness of the Triage framework in identifying and combating cybercrime related to online scams and 419 activities.                                                                                                                                                                                                                                                                                                                                                                                                                                                                                                                                                                                                                                                                     \\ 
\hline
Ephraim et al. \cite{ephraim2013african}       & This research investigates the effectiveness of a culture-centered approach in combatting cybercrime among African youths, focusing on West Africa. It highlights the need to address underexplored areas of cybercrime, such as cyberbullying and cyber violence, within the context of the rapidly growing internet penetration in the region.                                                                                                                                                                                                                                                                                                                                                                                                                                                                                                                                                                                                                                                                                                                 \\ 
\hline
David-West et al. \cite{David-West2021Sep}     & Challenges hindering the diffusion of mobile money in Nigeria were discussed in this paper, and long-term solutions are proposed to enhance financial inclusion and potentially reduce the risk of cybercrime in the country.                                                                                                                                                                                                                                                                                                                                                                                                                                                                                                                         \\ 
\hline
Abdulrauf et al. \cite{abdulrauf2017personal}  & The study emphasized the need for an updated regulatory framework to protect individuals' personal data and prevent potential violations as the country experiences rapid advancements in Information and Communication Technology.                                                                                                                                                                                                                                                                                                                                                                                                                                                                                                                                                                                                                                                                                                                                                                                                                        \\ 
\hline
Adeleke et al. \cite{Adeleke2022Jun}           & Spatial and socioeconomic inequalities in Financial Inclusion (FI) in Nigeria are the major drivers of FI. This paper examines how FI can inform region-specific policies in Nigeria and potentially reduce the incidence of cybercrime related to financial exclusion.                                                                                                                                                                                                                                                                                                                                                                                                                                                                                                                                                                                                                                                 \\ 
\hline
Dziwornu et al. \cite{dziwornu2021crime}       & The research investigates crime patterns and trends in Ghana, revealing a significant decline in crime rates. It assesses the potential influence of macro-level factors, situational crime prevention strategies, and the increasing presence of private security companies on the observed reduction in crime.                                                                                                                                                                                                                                                                                                                                                                                                                                                                                                                                                                                                          \\ 
\hline
Olubusoye et al. \cite{Olubusoye2023Apr}       & The research investigated the nature and causes of youth unemployment in Nigeria, which is a contributing factor to cybercrime in West Africa and proposes evidence-based solutions, including fiscal and monetary policy easing, to address the issue.                                                                                                                                                                                                                                                                                                                                                                                                                                                                                                                                                                                                                                                                  \\ 
\hline
Olanrewaju Lawal \cite{Lawal2022Nov}           & This paper explores how lessons from Nigeria's handling of the COVID-19 pandemic can be applied to managing the impacts of climate change, identifying gaps in management, and proposing practical approaches for resilience building, including the need for investment in preparedness, education, research, and data infrastructure.                                                                                                                                                                                                                                                                                                                                                                                                                                                                                                                                                                                      \\
\hline
\end{tabular}
\end{table*}

In exploring the second research question, table (\ref{summarytable}) presents a summarized overview of the seminal articles reviewed in this study. Notably, the study conducted by Ephraim et al.~\cite{ephraim2013african} investigates the implementation of a culture-centered approach in the context of cybercrime. This approach advocates for the incorporation of information ethics and the respect for individual dignity and rights, leveraging the distinct cultural values and ethical practices unique to each community. The study underscores the significance of cultural norms, values, and systems in shaping effective strategies against cybercrime. The effectiveness of cybercrime prevention in West Africa can be enhanced by promoting ethical online behavior grounded in the region's cultural values. By recognizing and respecting the diversity and dynamics of cultural norms, this strategy contributes to the development of tailored solutions that resonate with the local population. Ultimately, a culture-centered approach aims to foster a sense of responsibility and awareness among individuals in the digital realm, helping to create a safer online environment and minimize the prevalence of cybercrime in West Africa~\cite{ephraim2013african}.
Beek et al.~\cite{Beek2016May} argue that each case of cybercrime can be seen as a series of interconnected stories and that the narratives of scammers, victims, and police officers are shaped by cultural imaginaries and practices \cite{Beek2016May}.
The theoretical framework in the research of Kanu et al. \cite{kanu2022frauds} study addresses our Research Question 2 by highlighting the increase in cybercrime and the inadequacy of current countermeasures. To tackle this issue, the study proposes several techniques. One recommendation is to train staff on e-banking safety measures to enhance security. Additionally, the study suggests the adoption of blockchain technology for securing data transactions within financial institutions. By utilizing blockchain, fraud and forgery can be reduced through robust authentication and prevention of evidence tampering~\cite{kanu2022frauds}. Isacenkova et al. \cite{isacenkova2014inside} utilized the Triage approach for identifying scam campaigns. This approach utilizes a software framework for data mining that leverages multi-criteria data analysis to group events based on subsets of common elements. The Triage framework has been demonstrated to be effective in security investigations by employing a four-stage process: 1) feature selection, 2) graph-based clustering, 3) multi-criteria aggregation (Midek), and 4) cluster visualization \cite{thonnard2010multi}. This structured method helps researchers better understand and track the patterns of scam emails, ultimately aiding in the detection and prevention of such activities \cite{isacenkova2014inside}. 
Dziwornu et al.\cite{duah2015impact} conducted an in-depth analysis of macro-level crime trends in Ghana, utilizing crime data from 2000 to 2015, sourced from the Statistical and Information Technology Unit of the Criminal Investigations Department (SITU). The results demonstrated a considerable decline in crime rates throughout the three and a half-decades of investigation. The authors concluded that several factors, including national-level macroeconomic indicators like a robust economy, shifting demographics, and incarceration, contributed to reducing crime rates. The study's central question was to determine the effectiveness and responsibility of the comprehensive situational crime prevention strategy in contributing to the observed decrease in crime. Moreover, other researchers have noted the rising presence of private security companies in major Ghanaian cities, which may indicate an increasing public perception of crime \cite{owusu2016assessment}. In the context of cybercrime prevention in West Africa, this study implies that situational crime prevention strategies, along with other macro-level factors, could potentially effectively reduce cybercrime rates in the region \cite{dziwornu2021crime}.

\section{Discussion}
\subsection{Financial Crime}
The phrase "419" originates from Section 419 of the Nigerian penal code and encompasses a variety of scam activities, including fraudulent transactions, monetary theft, deceptive advertising, and more. These acts can be executed through various methods, such as social engineering, hacking, and brute force attacks. In order for victims to communicate with the perpetrators, contact information such as email addresses and phone numbers must be genuine, often remaining consistent and reused over an extended time period. A study by Isacenkova et al. \cite{isacenkova2014inside} revealed the existence of around 1,000 distinct scam strategies through their comprehensive analysis.
The study of Kanu et al. \cite{kanu2022frauds} explored current fraud and forgery cases in the Nigerian banking industry. The study showed that the incidence of these crimes increased by 40\% in 2019 from 2018 and 190\% in 2020 from 2019. In their study, the authors highlighted the growing issue of fraud in the banking industry, which includes activities such as stock market manipulation, money laundering, and more. Organizational fraud is increasing worldwide, especially in countries like Latin America, the Caribbean, India, Southeast Asia, and African countries where incidents of fraud and forgery are intensifying due to the growing use of electronic transactions impacting companies and financial institutions. The Central Bank of Nigeria reported 26,182 attempted frauds and forgeries in 2017, resulting in an approximate loss of N2.4 billion \cite{kanu2022frauds}.
Isacenkova et al. \cite{isacenkova2014inside} study identified four distinct scam campaigns:
\begin{enumerate}
    \item \textit{ESKOM Initiative:} This initiative is segmented into two distinct sub-initiatives. The initial phase is characterized by a fraudulent lottery scheme spanning one year, followed by a subsequent phase that extends for one and half years, involving the impersonation of ESKOM Holdings, a prominent electricity provider based in South Africa.
    
    \item \textit{Sify-Rolex Operation:} Central to this operation, identified as a 419 scam, were email addresses and contact numbers, persisting over a one and half year period with thematic shifts occurring every one to two months, totaling five changes. The continuity in phone number usage across all sub-operations suggests they were orchestrated by a singular scamming entity.
    
    \item \textit{iPhone campaign:} This campaign focused on fraudulent iPhone-related activities.
    \item \textit{Internationally-operated campaigns:} These exploit fake lottery topics, such as a Spanish fake lottery campaign that uses topic-related email addresses to appear more legitimate.
\end{enumerate}
The study discovered that some of these macro-campaigns spanned multiple countries, including both African and European nations. The infrastructure, orchestration, and modus operandi of these campaigns differed from traditional spam campaigns, extensively utilizing anonymization tools like webmail accounts to hide IPs and anonymous proxy phone numbers ~\cite{isacenkova2014inside}.
\subsection{Cybersecurity}
West Africa has a population of 367 million as of 2015 and is home to 5 percent of the global population, according to the United Nations (UN) \cite{USGS2016}. The region has been experiencing rapid population growth at an average annual rate of 2.75 percent, and urbanization has been progressing at a remarkable pace, with major cities recording mean annual growth rates of up to 9 percent. Fast forward to January 2022, Nigeria, a West African country, registered approximately 109 million active internet users, representing about half of the nation's total population. The country's internet penetration rate reached 51.0 percent at the beginning of 2022, with an estimated increase of 4.8 million users (+4.6\%) between 2021 and 2022 \cite{DataReportal,adewopo2021exploring}. This upward trend in cyberspace utilization not only demonstrates the growing digital landscape in Nigeria but also points to the potential rise in the use of cyberspace for criminal activities \cite{adewopo2020exploring}.
Abdulrauf et al. \cite{abdulrauf2017personal} research highlights the importance of personal data protection and the lack of an effective legal framework in Nigeria. Although the paper does not directly address cybercrime prevention methods in West Africa, it implies that the absence of a strong legal framework for data protection in Nigeria contributes to the overall ineffectiveness of cybercrime prevention efforts in the region. By suggesting appropriate legal reforms, the paper emphasizes the need for an updated regulatory framework to protect individuals' personal data and prevent potential violations as the country experiences rapid advancements in Information and Communication Technology.
The growth of international trade and digital commerce has been pivotal in the evolution of IT infrastructure within Nigeria. The Nigerian government is actively pursuing upgrades to its broadband services to bolster this infrastructure further. Enacting a data protection statute in Nigeria is a critical step in establishing a robust legal framework that will protect data residing in the databases of key public and private sector organizations \cite{abdulrauf2017personal}. In contrast, Ghana is recognized for its political stability and secure democratic practices within the Sub-Saharan African belt. While Ghana's crime rates are relatively modest compared to many developed nations, it still faces challenges in crime, which are a notable issue within its social fabric \cite{dziwornu2021crime}. Nigeria and Ghana can progress towards a fortified digital milieu for their populace by focusing on IT infrastructure enhancements and comprehensive data protection legislation. As these nations continue to experience growth and stability, it is essential to prioritize cybersecurity and data protection, ensuring the safety and well-being of their populations \cite{dziwornu2021crime}.
\subsection{Cybercrime}

The study of Olofinbiyi et al. \cite{olofinbiyi2021exploring} focused on analyzing the level of awareness among youth regarding cybercrime and probing into the causes that contribute to its widespread occurrence in Lagos Metropolis, Nigeria. The study discovered several contributing factors, including urbanization, unemployment and poverty, influence from peers, extensive knowledge in science and technology, inadequate political will to combat cybercrime, and ineffective enforcement of cybercrime laws.
The study also highlights two awareness theories;
According to the anomie theory, individuals who feel a sense of detachment from society or lack clear norms may be inclined to participate in deviant conduct, including cybercrime. In the realm of cybercrime, anomie theory proposes that the fast-paced technological advancements and the absence of well-defined regulations regarding online conduct may fuel the spread of cybercrime.
On the other hand, differential association theory argues that people learn deviant actions by interacting with others. In the context of cybercrime, this theory suggests that young individuals who are exposed to peers engaging in cybercrime are more likely to engage in such behavior themselves. Furthermore, the theory proposes that individuals who possess advanced knowledge in science and technology may be more prone to committing cybercrime due to their increased ability to exploit weaknesses in computer systems.


Hamisu et al. \cite{hamisu2021analysis} presents a comprehensive outline of cybercrime in Nigeria, including the commonly perpetrated types. The study also evaluates the initiatives of the Nigerian government in combating cybercrime and identifies areas of effectiveness and shortcomings. They later provide recommendations and suggestions for enhancing the effectiveness of law enforcement and government efforts to combat cybercrime in Nigeria.
The paper makes several recommendations to curb cybercrime in Nigeria, including:
\begin{enumerate}
    \item Strengthening the legal framework for addressing cybercrime by enacting and enforcing laws that criminalize cybercrime and provide for appropriate penalties.
    \item Improving the capacity of law enforcement agencies to investigate and prosecute cybercrime by providing them with the necessary training, equipment, and resources.
    \item Enhancing international cooperation and collaboration in the fight against cybercrime by working with other countries to share information and intelligence, and to extradite cyber criminals.
    \item Raising public awareness about the dangers of cybercrime and how to protect oneself from it by launching public education campaigns and providing cybersecurity training to individuals and organizations.
    \item Encouraging the development of a robust cybersecurity industry in Nigeria by providing incentives for cybersecurity startups and investing in research and development.
    \item Addressing the root causes of cybercrime, such as poverty and unemployment, by creating more job opportunities and improving the economic situation in the country.
\end{enumerate}
Overall, the paper suggests that a multi-faceted approach is needed to effectively tackle cybercrime in Nigeria, which involves not only law enforcement efforts but also addressing the underlying social and economic factors that contribute to the problem.

\section{Limitation}
This study only included articles published in English; some West African countries' official language is French and Portuguese, which may have resulted in the exclusion of relevant articles published in other languages. 
From our analysis, we noticed the geographical scope of the research study included mainly focused on papers from Nigeria, Ghana, and South Africa, with limited coverage of other West African countries. This limitation may have resulted in the exclusion of relevant studies from other West African countries.
Furthermore, some important papers may have been omitted because their keywords did not contain the phrase "West Africa." The authors may have generalized the research to include Africa as a whole, which may have caused some relevant studies to be excluded.
While this study provides valuable insights into the methods for preventing cybercrime in West Africa, its limitations suggest that there is a need for further research that addresses the above limitations and provides a more comprehensive understanding of the topic.
\section{Conclusion}
The rapid population growth, urbanization, and increasing internet penetration in West Africa, particularly in Nigeria, present both opportunities and challenges for the region. As more people gain access to the internet, there is a greater potential for economic growth and technological advancements \cite{DataReportal}. However, the expanding digital landscape also provides cyber criminals with more opportunities to exploit unsuspecting users, emphasizing the need for robust cybersecurity measures and public awareness campaigns to combat cybercrime effectively. Despite facing numerous political and economic challenges, Africa is experiencing a transformative shift due to globalization, which is evident in the growing industrialization, trade, education, and technological advancements across the continent. This global integration fostered development in Africa, elevating many African nations to  the status of developing countries \cite{ephraim2013african}. The rise of emerging economies, particularly in West African countries such as Nigeria, alongside other African countries like South Africa and Kenya, highlights the continent's progress and potential for further growth \cite{Fage2023,USGS2016}. 
Ephraim et al.~\cite{ephraim2013african} observed that current research predominantly concentrates on cybercrime in Africa, such as cyber fraud and cyber terrorism. However, it is essential to broaden the scope of research to encompass other critical areas, including cyberbullying and cyber violence. By exploring these additional dimensions of cybercrime, researchers can gain a more comprehensive understanding of the challenges faced by African individuals and communities and develop more effective prevention strategies and policies to combat these pressing issues.
The Triage framework, which was presented in \cite{isacenkova2014inside}, can be leveraged to combat cybercrime, especially in the context of online scamming and 419-related activities. This systematic approach helps researchers and authorities better understand and track the patterns of such scams, ultimately aiding in their detection and prevention \cite{isacenkova2014inside}.

\bibliographystyle{IEEEtran}
\bibliography{Main}

\end{document}